%% file: main.tex
\documentclass[runningheads]{llncs}
\pdfoutput=1
\usepackage{etex} 

\usepackage{xcolor}
\definecolor{amazing}{RGB}{254,67,101}

\include{header}

\usepackage{authblk}

\newcommand{\const}{\mathsf{c}}
\newcommand{\cproj}[1][r]{\ensuremath{\const^\text{proj}_{#1}}}

\renewcommand{\G}{\mathcal G}
\newcommand{\proj}{P}
\newcommand{\bekernel}{sparse kernel}  

\begin{document}

\title{Harmless Sets in Sparse Classes}
\author{
  P{\aa}l~Gr{\o}n{\aa}s~Drange\inst{1}\orcidID{0000-0001-7228-6640}
  \and
  Irene~Muzi\inst{2}\orcidID{0000-0003-2410-6523}
  \and
  Felix~Reidl\inst{3}\orcidID{0000-0002-2354-3003}
}
\institute{
  University of Bergen, Norway, \email{Pal.Drange@uib.no}
  \and
  \email{Irene.Muzi@gmail.com}
  \and
  Birkbeck College, University of London, United Kingdom, \email{felix.reidl@gmail.com}
}

\authorrunning{Drange, Muzi, Reidl}

\maketitle
\begin{abstract}
  In the classic \name{Target Set Selection} problem, we are asked to
  minimise the number of nodes to \emph{activate} so that, after the application
  of a certain propagation process, all nodes of the graph are active.
  Bazgan and Chopin
  [\textit{Discrete Optimization}, 14:170--182, 2014]
  introduced the opposite problem, named \name{Harmless
    Set}, in which they ask to maximise the number of nodes to activate
  such that not a single additional node is activated.

  In this paper we investigate how sparsity impacts the tractability of
  \name{Harmless Set}. Specifically, we answer two open questions posed
  by the aforementioned authors, namely a) whether the problem is \FPT{}
  on planar graphs and b) whether it is \FPT{} parametrised by treewidth.
  The first question can be answered in the positive using existing meta-theorems on  sparse classes, and we further show
  that \name{Harmless Set} not only admits a polynomial kernel, but that it can be solved in subexponential time.
  We then answer the second question in the negative
  by showing that the problem is W[1]-hard when
  parametrised by a parameter that upper bounds treewidth.
\end{abstract}


%
\section{Introduction}

How information and cascading events spread through social and
complex networks is an important measure of their underlying systems, and is a
well-researched area in network science.
The dynamic processes governing the diffusion of information and
``word-of-mouth'' effects have been studied in many fields, including
epidemiology, sociology, economics, and computer
science~\cite{kempe2003maximizing,kempe2005influential,
  easley2010networks, chen2009approximability}.

A classic propagation problem is the \name{Target Set Selection}
problem, first studied by Domingos and
Richardson~\cite{domingos2001mining, richardson2002mining}, and later
formalised in the context of graph theory by
Chen~\cite{chen2009approximability, chen2009efficient}.
Chen defines the problem as how to find $k$ initial \emph{seed} vertices
that when activated cascade to a maximum; this model is called
\emph{standard independent cascade model of network diffusion}.
It has also been studied under the name of \name{Influence
  Maximization}~\cite{nguyen2012containment,nguyen2016stop} in the
context of lies spreading through a network~\cite{chen2011influence,
  chen2013information}, bio-terrorism~\cite{dreyer2009irreversible}, and
the spread of fires~\cite{roberts2006graph}.
Information propagation is modelled as an \emph{activation process}
where each individual is activated if a sufficient number of its
neighbours are active. Sufficient here means that the number of active
neighbours of an individual $v$ exceeds a given threshold $t(v)$ which
is assigned to each individual to capture their resilience to being
influenced.

Motivated by \emph{cascading of information} we study vertices that are
\emph{harmless}, i.e., a set of vertices that can be activated without
any cascades whatsoever.
However, activating all vertices in a graph is a trivial solution in the
standard diffusion model, since we cannot cascade further.
We therefore want to differentiate between \emph{initially activated
  vertices} and vertices that have been \emph{activated by a cascade}.
In this setting, we can therefore say that we want a largest possible
set of initially activated vertices that do not cascade at all, even to
itself.
It was first studied by Bazgan and Chopin \cite{HSParameterized14} under the name \HS, who showed that it is
\W[2]-complete in general and \W[1]-complete if thresholds are bounded
by a constant. They observe (see Observation~\ref{obs:k-bounded} below)
that one can bound the maximum threshold by the solution size
and thus obtain a simple \FPT{} algorithm when parametrised by
the solution size $k$ \emph{and} the treewidth.
Bazgan and Chopin conclude their work with the following open questions:

\begin{open*}[Bazgan and Chopin~\cite{HSParameterized14}]
  Is \HS{} fixed-parameter tractable on
  \begin{enumerate}
  \item general graphs with respect to the parameter treewidth?
  \item on planar graphs with respect to the solution size?
  \end{enumerate}
\end{open*}
\noindent
Here we answer both these problems: no and yes, and simultaneously
discover surprising connections between \HS{} and \name{Dominating Set}
in sparse graphs.

\bigskip

\paragraph{Our results.}

Let us distinguish two flavours of this problem: \name{$p$-Bounded Harmless
  Set}, where we consider the bound~$p$ a constant, and
%
\name{Harmless Set}
where the threshold is unbounded.

\begin{XProblem}{Harmless Set}
  \Input & A graph $G$ with a threshold function $t\colon V(G) \to \mathbb N_{> 0}$
           and an integer $k$. \\
  \Prob & Is there a vertex set $S \subseteq V(G)$ of size at least $k$ such
          that every vertex $v \in G$ has fewer than $t(v)$ neighbours in $S$?
\end{XProblem}

\begin{XProblem}{$p$-Bounded Harmless Set}
  \Input & A graph $G$ with a threshold function $t\colon V(G) \to \mathbb [p]$
           and an integer $k$. \\
  \Prob & Is there a vertex set $S \subseteq V(G)$ of size at least $k$ such
          that every vertex $v \in G$ has fewer than $t(v)$ neighbours in $S$?
\end{XProblem}


\noindent
Note that harmless sets are
hereditary in the sense that if $S$ is a harmless set of an instance $(G,t)$, then any subset $S' \subseteq S$ is also harmless for $(G,t)$.
Therefore instead of searching for a harmless set of size at least $k$, we
can equivalently search for a harmless set of size exactly $k$. In this scenario
we can replace all thresholds above $k$ with $k+1$:

\begin{observation}\label{obs:k-bounded}
  \HS{} parametrised by $k$ is equivalent to
  \name{$(k+1)$-Bounded Harmless Set} parametrised by $k$.
\end{observation}

\noindent
Let us begin by briefly answering the first question of Bazgan and
Chopin in the positive.  It turns out that a simple of application the
powerful machinery of first-order model checking%
\footnote{
    There exist some intricacies regarding the type of nowhere dense class
    and whether the resulting \FPT{} algorithm is uniform or not. This is just
    a technicality in our context and we refer the reader to Remark 3.2 in~\cite{NowhereDenseFO17}
    for details.
  }
in sparse classes~\cite{NowhereDenseFO17}
is enough (see Appendix for a short proof):

\begin{restatable}{proposition}{fptnowheredense}
  \HS{} parametrised by $k$ is fixed-parameter tractable in
  nowhere dense classes.
\end{restatable}

\noindent
We will briefly discuss the notion of nowhere denseness below, in this context
it is only important that planar graphs are nowhere dense.

These previous results and our observation regarding tractability in sparse classes
leave two important questions for us. First, does the problem admit a polynomial
kernel in sparse classes? And second, is there a chance that the problem could be
solved on \eg graphs of bounded treewidth without parametrising by the
solution size? In the following we answer the kernelization question in the affirmative:

\begin{theorem}
 \HS{} admits a polynomial \bekernel{} in classes of bounded expansion. \name{$p$-Bounded Harmless Set}, for any constant $p$, admits a
 linear \bekernel{} in these classes.
\end{theorem}

\noindent
Classes with bounded expansion include planar
graphs (and generally graphs of bounded genus),
graphs of bounded degree, classes excluding a (topological) minor, and more.
The term \emph{\bekernel} is explained below in
Section~\ref{sec:be-classes}; It alludes to the fact that the
constructed kernel does not necessarily belong to the original graph
class but is guaranteed to be ``almost'' as sparse.

Bazgan and Chopin give an algorithm for \name{Harmless set} parametrised
by treewidth and the solution size running in time
$O(k^{O(\text{tw})} \cdot n)$, when provided a tree decomposition as
part of the input%
\footnote{This can be relaxed using a constant factor, linear time
  approximation for computing tree
  decompositions~\cite{bodlaender2016approximation}}%
.  They conclude by asking whether the problem is ``fixed-parameter
tractable on general graphs with respect to the parameter treewidth
[alone]''~\cite{HSParameterized14}.  We answer this question in the
negative:

\begin{restatable}{theorem}{lowerbound}\label{thm:lowerbound}
  \HS{} is \emph{\W[1]}-hard when parametrised by a modulator
  to a $2$-spider-forest\footnote{A $1$-spider-forest is a starforest, and a $2$-spider-forest is a subdivided starforest}.
\end{restatable}

\noindent
Since a $2$-spider-forest has treedepth, pathwidth, and treewidth at
most 3, a graph with a modulator $M$ to a $2$-spider-forest has
treedepth, pathwidth, and treewidth at most $|M| + 3$.
This very strong structural parametrisation means that the problem is
not only hard on general sparse graphs, but indeed also \W[1]-hard for
parameters like treewidth, pathwidth, and even treedepth.
We complement this result by showing that a slightly stronger parameter,
the vertex cover number, does indeed make the problem tractable:

\begin{restatable}{theorem}{vcalgorithm}\label{thm:vcalgorithm}
  \HS{} is fixed-parameter tractable when parametrised
  by the vertex cover number of the input graph.
\end{restatable}

\paragraph{Note.}
We obtained our results simultaneously with and
independent from those by Gaikwad and
Maity~\cite{gaikwad2021harmless}.  They provide an explicit and potentially
practical FPT algorithm for planar graphs while we show that the problem is 
not only FPT on planar graphs, but indeed on a much more general class of graphs, namely those of bounded expansion.  We also show that on
apex-minor-free graphs (which include planar graphs),
there exists a subexponential time algorithm for the problem.  That is,
we show the following results, which improves on Gaikwad and Maity's
$2^{O(k \log k)} n^{O(1)}$ algorithm for planar graphs:

\begin{restatable}{theorem}{subexpalg}\label{thm:subexpalg}
  \HS{} is solvable in time $O( 2^{o(k)} \cdot n )$ on apex-minor-free
  graphs.
\end{restatable}

\section{Preliminaries}

\marginnote{$|G|$, $\|G\|$, $2$-spider}
For a graph $G$ we use $V(G)$ and $E(G)$ to refer to its vertex- and edge-set,
respectively. We used the short hands $|G| := |V(G)|$ and $\|G\| := |E(G)|$.
A $2$-spider is a graph obtained from a star by subdividing every edge at most once. A $2$-spider-forest is the disjoint union of arbitrarily many $2$-spiders.

\marginnote{$f(G)$, $f(X)$, $N(X)$, $N^r(\any)$, {$N^r[\any]$}}
For functions $f\colon V(G) \to \mathbb R$ we will often use the shorthands
$f(X) := \sum_{u \in X} f(x)$ and $f(G) := f(V(G))$. Similarly, we use the
shorthand $N(X) := (\bigcup_{u \in X} N(u)) \setminus X$ for all neighbours of
a vertex set~$X$. The $r$\th neighbourhood $N^r(u)$ contains all vertices at distance
exactly~$r$ from~$u$, the closed $r$\th neighbourhood $N^r[u]$ all vertices at
distance at most~$r$ from~$u$ (also known as the $r$-ball of $u$).
This corresponds to $N(u) = N^1(u)$ and $N[u] = N^1[u]$.
We refer to the textbook by Diestel~\cite{diestel2016graph} for more on
graph theory notation.

\marginnote{$r$-scattered, $r$-dominating, $\ds_r(G)$, $\ds_r(G,X)$}
A vertex set $X \subseteq V(G)$ is \emph{$r$-scattered} if
for $x_1 \in X$ and $x_2 \in X$, $N^r[x_1] \cap N^r[x_2] = \emptyset$.
Equivalently, $N^r[u] \cap X \leq 1$ for \emph{all} vertices $u \in G$,
or the pairwise distance between members of $X$ is at least $2r+1$.
A vertex set $D \subseteq V(G)$ is
\emph{$r$-dominating} if $N^r[D] = V(G)$ and we write $\ds_r(G)$ to denote
the minimum size of such a set. Similarly, we say that $D$ $r$-dominates another
vertex set $X \subseteq V(G)$ if $X \subseteq N^r[D]$ and we write $\ds_r(G,X)$
for the minimum size of such a set. In both cases we will omit the subscript $r$
for the case of $r = 1$. In classes with bounded expansion, the size of
$r$-scattered sets is closely related to the $r$-domination number, see the
toolkit section below.

Given a vertex set $X \subseteq V(G)$ we call a path
\marginnote{$X$-avoiding, $r$-projection}
\emph{$X$-avoiding} if its internal vertices are not contained
in $X$. A \emph{shortest $X$-avoiding path} between vertices $x,y$
is shortest among all $X$-avoiding paths between $x$ and $y$.

\begin{definition}[$r$-projection]
  For a vertex set $X \subseteq V(G)$ and a vertex $u \not \in X$
  we define the \emph{$r$-projection} of $u$ onto $X$ as the set
  \vspace*{-6pt}
  \[
    \proj^r_X(u) := \{ v \in X \mid \text{there exists an $X$-avoiding $u$-$v$-path of length} \leq r \}
    \vspace*{-6pt}
  \]%
\end{definition}

\noindent
Two vertices with the same $r$-projection onto $X$ do not, however, necessarily
have the same (short) distances to $X$. To distinguish such cases, it is useful
to consider the \emph{projection profile} of a vertex to its projection:

\begin{definition}[$r$-projection profile]
  For a vertex set $X \subseteq V(G)$ and a vertex $u \not \in X$
  we define the \emph{$r$-projection profile} of $u$ onto $X$ as a function
  $\pi^r_{G,X}[u] \colon X \to [r] \cup \infty$ where $\pi^r_{G,X}[u](v)$
  for $v \in X$ is the length of a shortest $X$-avoiding path from $u$ to $v$
  if such a path of length at most $r$ exists and $\infty$ otherwise.
\end{definition}

\subsection{Bounded expansion classes and kernels}\label{sec:be-classes}

\Nesetril and Ossona de Mendez~\cite{BndExpI} introduced bounded expansion
as a generalisation of many well-known sparse classes like planar graphs, graphs
of bounded genus, bounded-degree graphs, classes excluding a (topological)
minor, and more. The original definition of bounded expansion classes made use of the
concept of \emph{shallow minors} inspired by the work of Plotkin, Rao, and
Smith~\cite{PlotkinRaoSmith94}.

\begin{definition}
  A graph $H$ is an \emph{$r$-shallow minor} of $G$, written as
  $H \sminor^r G$, if $H$ can be obtained from
  $G$ by contracting disjoint sets of radius at most~$r$.
\end{definition}

\noindent
Classes of bounded expansion are then defined as those classes in which the
density (or average degree) of $r$-shallow minors is bounded by a function of~$r$.

\begin{definition}\marginnote{grad, $\grad_r(\any)$}
  The \emph{greatest-reduced average degree} (grad) $\grad_r$ of a graph~$G$
  is defined as
  \[
    \grad_r(G) = \sup_{H \sminor^r G} \frac{\|H\|}{|H|}
  \]
\end{definition}

\begin{definition}
  A graph class $\G$ has \emph{bounded expansion} if there exists a function~$f$
  such that $\grad_r(G) \leq f(r)$ for all $G \in \G$.
\end{definition}

\noindent
For example, it is easy to see that classes with maximum degree $\Delta$ have
bounded expansion with $f(r) := \Delta^{r+1}$. In the following we will often make use of the property that the grad of a graph does not change much under the addition of a few high-degree vertices: if $G$ is
a graph and $G'$ is obtained from $G$ by adding an apex-vertex, then $\grad_r(G')
\leq \grad_r(G) + 1$.

One principal issue with designing kernels for bounded expansion classes is the uncertainty
of whether certain gadget constructions preserve the class or not. When working
with more concrete classes like planar graphs we can be certain that \eg adding
pendant vertices will result in a planar graph.
When working with some
arbitrary bounded expansion class $\G$ this is not necessarily possible: $\G$ might,
for example, consist of all graphs with grad bounded by some function \emph{and} minimum degree at least two. In such cases, the addition of a pendant vertex takes us outside of the class even though the grad did not increase.

\marginnote{\bekernel}
We resolve this issue as proposed in the paper~\cite{Waterlilies}. Let
$\Pi$ be a parametrised problem over graphs. A \emph{\bekernel} of $\Pi$
is a kernelization for which there exists a function $g$ that, given an
instance with graph~$G$, outputs a graph~$G'$ that besides the usual
constraints on the size~$|G|+\|G'\|$ further satisfies that
$\grad_r(G') \leq g(\grad_r(G))$ for all $r \in \N$. Therefore if the input
graphs are taken from a bounded expansion class~$\cal G$, the outputs will also belong to a,
potentially different, bounded expansion class~$\cal G'$.

\subsection{The bounded expansion toolkit}

\noindent
The notion of independence (or more specifically scatteredness) plays a
central role in the theory of sparse graphs. As a prime example,
Dawar~\cite{dawar2007finite,dawar2010homomorphism} introduced the
notions of wideness and quasi-wideness---both related to independence---as one possible classification of
sparseness.
We will need the following definition from his work;
recall that an $r$-scattered set is a set of vertices $X \subseteq V(G)$
such that for any vertex $u$ in $V(G)$, the $r$-ball of $u$ contains at
most 1 vertex from $X$.

\begin{definition}\marginnote{uniformly quasi-wide}
  A class $\mathcal G$ is \emph{uniformly quasi-wide} if for every $m \in
  \mathbb N$ and $r \in \mathbb N$ there exist numbers $N = N(m,r)$ and $s
  =s(r)$ such that the following holds:

  Let $G \in \mathcal G$ and let $A \subseteq V(G)$ with $|A| \geq N$. Then
  there exists $S \subseteq V(G)$, $|S| \leq s(r)$ and a set $B \subseteq A -
  S$, $|B| \geq m$, such that $B$ is $r$-scattered in $G-S$.
\end{definition}

\noindent
As it turns out this notion of sparseness coincides with the notion of
\emph{nowhere denseness} in graph classes closed under taking subgraphs~\cite{FOPropertiesND10}.
Bounded expansion classes are nowhere dense and the following result due to
Kreutzer, Rabinovich, and Siebertz plays a crucial role in our kernelization
procedure.

\begin{theorem}[Kreutzer, Rabinovich, Siebertz~\cite{DSKernelND}]\label{thm:uqw}
  Every nowhere dense class $\mathcal G$ is \emph{uniformly quasi-wide} with
  $N(m,r) = m^{g(r)}$ for some function~$g$. Moreover, there exists an algorithm which,
  given $G \in \mathcal G$ and $A \subseteq V(G)$ as input, computes an
  $r$-scattered set of the promised size in time $|A|^{O(1)} n^{1+o(1)}$.
\end{theorem}

\noindent
There is a second method to compute suitable scattered sets which we can
leverage to create a ``win-win'' argument for our kernelization procedure. Concretely,
\Dvorak's algorithm~\cite{DvorakDomset} provides us either with a small $r$-dominating set or
a large $r$-scattered set. The following variant of the original algorithm
is called the \emph{warm-start} variant~(see \eg \cite{Waterlilies}):

\begin{theorem}[\Dvorak's algorithm~\cite{DvorakDomset}]\label{thm:dvorak-ds}
  For every bounded expansion class $\G$ and $r \in \N$ there exists
  a polynomial-time algorithm that, given a vertex set $X \subseteq V(G)$,
  computes an $r$-dominating set $D$ of $X$
  and an $r$-scattered set $I \subseteq D \cap X$ with $|D| = O(|I|)$.
\end{theorem}

\noindent
Note that since an $r$-scattered set $I \subseteq X$ provides a lower bound
for the $r$-domination of $X$ we have that $|D| = O(\ds_r(G,X))$.

Finally, we will need the following two fundamental properties of bounded expansion classes. The first is a refinement on the neighbourhood complexity
characterisation of bounded expansion classes~\cite{NComplexity19}:

\begin{lemma}[Adapted from~\cite{DSKernel,DSKernelND}]\label{lemma:projbound}
  For every bounded expansion class $\G$ and $r \in \N$ there exists a constant $\cproj$
  such that for every $G \in \G$ and $X \subseteq V(G)$, the number of
  $r$-projection profiles realised on $X$ is at most $\cproj |X|$.
\end{lemma}

\noindent
The second can be seen as a strengthening of the first: not only are the number of
projection profiles bounded linearly in the size of the target set, we can find a suitable superset of the target set which even restricts the \emph{size} of the projections to a constant.

\begin{lemma}[Projection closure~\cite{DSKernel}]\label{lemma:projclos}
  For every bounded expansion class $\G$ and $r \in \N$ there
  a polynomial-time algorithm that, given $G \in \G$ and $X \subseteq V(G)$,
  computes a superset $X' \supseteq X$, $|X'| = O(|X|)$, such that
  $|\proj^r_{X'}(u)| = O(1)$ for all $u \in V(G)\setminus X'$.
\end{lemma}

\subsection{Waterlilies}
Reidl and Einarson introduced the notion of \emph{waterlilies} as a structure
which is very useful in constructing kernels~\cite{Waterlilies}. We simplify the definition here as we do not need it in its full generality.

\begin{definition}[Waterlily]
  A \emph{waterlily} of \emph{radius~$r$} and \emph{depth~$d \leq r$}
  in a graph~$G$ is a pair
  $(R, C)$ of disjoint vertex sets with the following properties:
  \begin{itemize}
    \item $C$ is $r$-scattered in $G-R$,
    \item $N^r_{G-R}[C]$ is $d$-dominated by $R$ in $G$.
  \end{itemize}
  We call $R$ the \emph{roots}, $C$ the \emph{centres}, and the sets $\{N^r_{G-R}[x]\}_{x \in C}$
  the \emph{pads} of the waterlily.
  A waterlily is \emph{uniform} if all centres have the same
  $d$-projection onto $R$, \eg $\pi^d_R[x]$ is the same function for
  all $x \in C$.
\end{definition}

\noindent
We will frequently talk about the \emph{ratio} of a waterlily which we
define as a guaranteed lower bound of $|C|$ in terms of $|R|$,
\eg a waterlily of \emph{ratio $2|R| + 1$} satisfies $|C| \geq 2|R| + 1$.
The authors in~\cite{Waterlilies} used waterlilies with a constant ratio,
but a slight modification of their proof (in particular using
Theorem~\ref{thm:uqw}) lets us improve this ratio to any polynomial. We provide a proof with the necessary modification in the Appendix.

\begin{restatable}{lemma}{lilylemma}\label{lemma:lily}
  For every bounded expansion class $\G$ and $r,d \in \N$, $d \leq r$, the following
  holds. There exists a polynomial~$p_r$ such that for every
  $G \in \G$, $t \in \N$ and $A \subseteq V(G)$
  with $|A| \geq p_r(t) \ds_d(G,A)$ there
  exists a uniform waterlily $(R, C \subseteq A)$ with depth~$d$, radius~$r$,
  and with $|R| = O(1)$ and $|C| \geq t$,
  moreover, such a waterlily can be computed in polynomial time.
\end{restatable}

\section{A \bekernel{} for \pBHS}

In order to give a \bekernel{} we first show how to construct a bikernel
into the following annotated problem.

\begin{XProblem}{Annotated $p$-Bounded Harmless Set}
  \Input & A graph $G$ with a threshold function $t\colon V(G) \to \mathbb [p]$,
           an integer $k$, and a subset $K \subseteq V(G)$. \\
  \Prob & Is there a vertex set $S \subseteq K$ of size at least $k$ such
          that every vertex $v \in G$ has fewer than $t(v)$ neighbours in $S$?
\end{XProblem}

\marginnote{solution core}
\noindent
We call the set $K$ the \emph{solution core} of the instance (see~\cite{Waterlilies} for a general definition).
Next we present two lemmas whose application will step-wise construct smaller
annotated instances. The first lemma lets us reduce the size
of the solution core, the second the size of the graph. Afterwards we demonstrate
how these two reduction rules serves to construct a bikernel.

\marginnote{fragile}%
In the following, we often need to treat vertices with a threshold equal to one differently.
For brevity, we will call these vertices \emph{fragile}; observe that a fragile
vertex can be part of a solution but none of its neighbours can.
\begin{lemma}\label{lemma:shrink-core}
  Let $(G,t,k,K)$ be an instance of \name{Annotated $p$-Bounded Harmless Set}
  where $G$ is taken from a bounded expansion class and $K$ is a solution core.
  There exists a polynomial $q(p)$ such that
  the following holds: If $|K| \geq q(p) \cdot k$, then in polynomial time we either find that
  $(G,t,k,K)$ is a YES-instance or we identify a vertex $x \in K$ such
  that $K \setminus \{x\}$ is a solution core.
\end{lemma}
\begin{proof}
  First consider the case that there is a vertex $x \in K$ with a fragile
  neighbour $u \in N(x)$. Then $x$ of course cannot be in any solution
  and $K \setminus \{x\}$ is a solution core.

  Assume now that no vertex in $K$ has a fragile neighbour. We now use \Dvorak's
  algorithm (Theorem~\ref{thm:dvorak-ds}) to compute a $1$-dominating set for
  $K$; let $D$ be the resulting dominating set and $I \subseteq D \cap K$ the
  promised $1$-scattered set, i.e., with $|I| = \Omega(|D|)$.
  Since the neighbourhoods of vertices in $I$ are
  pairwise disjoint and no vertex in $I$ (as $I \subseteq K$) has a fragile
  neighbour, it follows that $I$ itself is a harmless set. So if $|I| \geq k$ we
  conclude that $(G,t,k,K)$ is a YES-instance.

  Otherwise $|I| < k$ and therefore, by Theorem~\ref{thm:dvorak-ds},
  $\ds(G,K) = O(k)$. We apply Lemma~\ref{lemma:lily} to compute a waterlily
  for the set $K$ at depth~$1$ and with radius~$2$. We will later choose $q(k)$
  to ensure that the following arguments go through.

  Let $(R,C \subseteq K)$ be the resulting uniform waterlily with
  $|C| \geq \kappa$, where $\kappa$ is an appropriately large value that
  we choose later.
  For the centres $v \in C$, define the following signature $\sigma(v)$:
  \[
    \sigma(v) = \{ (t(u), N(u) \cap R) \mid u \in N_{G-R}(v) \}.
  \]
  That is, $\sigma(v)$ records how neighbours of $v$ connect to $R$ and what
  thresholds these neighbours have. Define the equivalence relation~$\sim_\sigma$
  over $C$ via $v \sim_\sigma w$ iff $\sigma(v) = \sigma(w)$. Recall that, by
  Lemma~\ref{lemma:projbound} the number of $1$-projections onto $R$ is at most
  $\cproj[1]|R|$. Therefore we can picture $\sigma(v)$ as a string of length at most
  $\cproj[1]|R|$ over the alphabet $\{0,\ldots,p\}$ where $0$ indicates that a
  certain neighbourhood is not contained in $\sigma(v)$ and any non-zero value
  $a \in [p]$ indicates that this neighbourhood is realised by one of $v$'s
  neighbours with weight $a$.
  Accordingly, we can bound the index of $\sim_\sigma$ by
  \[
    | C / \sim_\sigma | \leq (p+1)^{\cproj[1]|R|}
  \]
  and thus by averaging there exists an equivalence class $C' \in C / \sim_\sigma$
  of size at least $|C| / (p+1)^{\cproj[1]|R|}$.

  We choose $|C|$ big enough so that $|C'| > (p-1)|R|$ and now claim that any
  vertex of $C'$ can be safely removed from $C$. To see this, fix an arbitrary
  vertex $x \in C'$. Consider any harmless set $S \subseteq K$ of size $k$, if
  no such set exists then $K \setminus \{x\}$ trivially is also a solution core.
  Note that if $x \not \in S$ we are done, so assume $x \in S$.

  \begin{claim}
    There exists a centre $x' \in C'$ such that $N^2_{G-R}[x']$, the pad
    of $x'$ in $(R,C')$, does not intersect $S$.
  \end{claim}
  \begin{proof}[Proof of claim]
    If $|N(R) \cap S| > (p-1)|R|$, there would be at least one vertex in $R$ whose
    threshold is exceeded, contradicting our assumption that $S$ is a harmless
    set. Since $R$ dominates the pads of $(R,C')$, they are all contained in
    $N(R)$ and we conclude that $S$ intersect at most $(p-1)|R|$ pads.
    Since $|C'| > p|R|$, the claimed centre $x' \in C'$ must exist.
  \end{proof}

  \noindent
  We claim that $S' := (S \setminus \{x\}) \cup x'$ is a harmless set. Note that $x \neq x'$
  since $x \in S$ but $x' \not \in S$, therefore $|S'| = |S| \geq k$. To show
  that $S'$ is harmless we show that no threshold of $N(x')$ is exceeded. This
  suffices since these are the only vertices whose threshold increases when
  $x$ is exchanged for $x'$. Fix $y' \in N(x')$ and consider the following cases.

  First, assume $y' \in R$. As $(R,C')$ is uniform, we have that
  $N(x') \cap R = N(x) \cap R$ and therefore $N(y') \cap S = N(y') \cap S'$.
  We conclude that $|N(y') \cap S'| = |N(y') \cap S| < t(y')$.

  Second, assume $y' \not \in R$. Since $\sigma(x) = \sigma(x')$, there exists
  a vertex $y \in N(x) \setminus R$ such that $t(y) = t(y')$ and $N(y) \cap R = N(y')
  \cap R$. Since $N^2_{G-R}[x']$, the pad of~$x'$, does not intersect~$S$
  and because $y' \in N_{G-R}(x')$ we have that $N(y) \cap S' = x' \cup (S' \cap R)$.
  Finally note that $|N(y) \cap S| < t(y)$ as $S$ is harmless.
  Therefore
  \[
    |N(y') \cap S'| =  1 + |S' \cap R| = 1 + |S \cap R| \leq |N(y) \cap S| < t(y).
  \]
  Since $t(y) = t(y')$, we conclude that $|N(y') \cap S'| < t(y')$.

  It follows that $S'$ is indeed a harmless set of size $|S|$ with $S' \subseteq
  K \setminus \{x\}$ and due to this exchange argument we find that $K \setminus \{x\}$ is indeed still
  a solution core for $(G,t)$. It remains to choose an appropriate polynomial
  $q(k)$. In the above arguments, we needed that $|C'| > (p-1)|R|$ which we now
  use to determine $q(p)$:
  \begin{align*}
    |C'| > (p-1)|R| & \implies |C| > (p+1)^{\cproj[1]|R|} (p-1)|R| \\
                    & \implies |K| > p_1 \left( (p+1)^{\cproj[1]|R|} (p-1)|R| \right) \cdot \ds(G,K)
  \end{align*}
  Where $p_1$ is the polynomial from Lemma~\ref{lemma:lily} (with $r=1$).
  Since by our very first argument $\ds(G,K) = O(k)$, it is therefore enough
  that $|K| > q(p) \cdot k$ with $q(p) = O(p_1((p+1)^{\cproj[1]|R|} (p-1)|R|)$.
  Since $|R| = O(1)$, $q$ is indeed a polynomial.
  Finally, note that all algorithmic steps (\Dvorak's algorithm, construction
  of the waterlily $(R,C')$) can be done in polynomial time.
\end{proof}

\noindent
The constant $\cproj[1]$ in the following lemma is the constant from
Lemma~\ref{lemma:projbound} for $r=1$.

\begin{lemma}\label{lemma:shrink-graph}
  Let $(G,t,k,K)$ be an instance of
  our \name{Annotated $p$-Bounded Harmless Set} problem
  where $G$ is taken from a bounded expansion class. Then, if
  $|K| < |G| / (\cproj[1]+1) $, then there exists a vertex $v \in V(G)\setminus K$
  such that $(G-v, t|_{V(G)-v},k,K)$ is an equivalent instance.
\end{lemma}
\begin{proof}
  Let $O := V(G)\setminus K$ for convenience.
  By Lemma~\ref{lemma:projbound}, the number of 1-projections that $O$
  realises on $K$ is bounded by $\cproj[1] |K|$. Accordingly, if $|O| >
  \cproj[1] |K|$, there exist two distinct vertices $u,v \in O$ with
  $\pi^1_K[u] = \pi^1_K[v]$ or, equivalently, $N(u) \cap K = N(v) \cap K$.
  Let wlog $t(v) \leq t(u)$, we claim that we can safely remove $v$ from
  the instance. To see this, consider a harmless set~$S \subseteq K$.
  Clearly, neither $u$ nor $v$ are in $S$. Furthermore, $N(u) \cap S = N(v) \cap S$,
  and since $|N(v) \cap S| \leq t(v)$ we also have $|N(u) \cap S| \leq t(u)$.
  We conclude that it is safe to remove $v$.
\end{proof}

\noindent
With these two reduction rules in hand, we can finally prove the main
result of this section.

\begin{theorem}\label{thm:bikernel}
  \name{$p$-Bounded Harmless Set} over bounded expansion classes admits a bikernel into
  \name{Annotated $p$-Bounded Harmless Set} of size $f(p) \cdot k$,
  for some polynomial~$f$.
\end{theorem}
\begin{proof}
  Let $I = (G,t,k)$ be an instance of \name{$p$-Bounded Harmless Set}.
  We first construct the instance $\hat I = (G,t,k,K)$ for
  \name{Annotated $p$-Bounded Harmless Set} where
  $K := \{ u \in G \mid \min_{v \in N(u)} t(v) > 1 \}$, that is, $K$ contains all
  vertices who do \emph{not} have a fragile neighbour.
  Observe that any solution $S$ for $I$ must necessarily avoid picking such vertices,
  and therefore $S \subseteq K$. We conclude that $I$ and $\hat I$ are
  equivalent instances.

  Now apply Lemma~\ref{lemma:shrink-core} iteratively to $\hat I$, that is, we
  apply the lemma until it either tells us that the current instance is a
  trivial YES-instance, in which case we output a constant-sized YES-instance
  and are done, or we arrive at an instance $\hat I' = (G,t,k,K')$ where $|K'|
  \leq q(p) \cdot k$.

  Next, we apply Lemma~\ref{lemma:shrink-graph} exhaustively to $\hat I'$, meaning
  we iteratively remove suitable vertices $v \not \in K$ until the resulting graph
  $G'$ satisfies $|G'| \leq (\cproj[1]+1)|K'|$. Call the resulting instance
  $\hat I'' = (G', t|_{V(G''), k, K'})$. By the bounds on $K'$ and $G'$ we have
  that
  \[
    |G'| \leq (\cproj[1]+1)|K'| \leq (\cproj[1]+1) q(p) \cdot k := f(p) \cdot k,
  \]
  as claimed.
\end{proof}

\begin{corollary}
  \HS{} admits a polynomial \bekernel{}.
\end{corollary}
\begin{proof}
  Given an instance $I = (G,t,k)$ of \HS{}, we first create the
  instance $\tilde I = (G, \tilde t, k)$ where $\tilde t$ is $t$ with thresholds
  larger than $k+1$ replaced by $k+1$. This, as we observed before, is an
  equivalent instance of \name{$(k+1)$-Bounded Harmless Set} and by
  Theorem~\ref{thm:bikernel} we can obtain a bikernel instance
  $\hat I = (\hat G, \hat t, k, \hat K)$ of size $f(k+1) \cdot k = k^{O(1)}$.

  We reduce back to \HS{} by constructing an instance $I' = (G', t', k)$
  from $\hat I$ as follows. Create $G'$ from $\hat G$ by adding two vertices
  $a, b$ where $a$ is connected to all of $K$ in $G'$ and $b$ is only connected to
  $a$. Set the thresholds $t'(a) = t'(b) = 1$ and let $t'$ be otherwise like $t$.
  To see that the two instances are equivalent, simply note that since $a$ and
  $b$ are fragile, no vertex of $N(a) \cup N(b) = V(G')\setminus K$ can be
  part of a harmless set. In other words, any solution of $I'$ must completely
  reside in $K$. The size of $I'$ differs to that of $I$ only by some constant
  factor, therefore we conclude that $I'$ is indeed a polynomial kernel of
  $I$. The construction itself increases the grad of $G$
  only by an additive constant (see Section~\ref{sec:be-classes}) therefore
  $I$ is indeed a \bekernel.
\end{proof}

\noindent
By the same construction we also obtain the following result:

\begin{corollary}
  \name{$p$-Bounded Harmless Set} for any constant~$p$ admits a
  linear \bekernel.
\end{corollary}

\section{Sparse parametrisation}

In this section we first prove Theorems~\ref{thm:lowerbound} and~\ref{thm:vcalgorithm}, namely that \HS{} is intractable
when parametrised by the size of a modulator to a \emph{$2$-spider-forest} but is
\FPT{} when parametrised by the vertex cover number of the input graph. We then
show that a simple application of the bidimensionality framework~\cite{demaine2008bidimensionality,fomin2009contraction} proves
Theorem~\ref{thm:subexpalg}, \ie that \HS{} can be solved in subexponential \FPT{} time on graphs excluding an apex-minor.

\subsection{Vertex cover}

\vcalgorithm*
\begin{proof}
  Let $(G,t)$ be an instance of \HS{} and let $X \subseteq V(G)$
  be a vertex cover of size~$2\vc(G)$ which we compute greedily by the usual local ratio algorithm. Let $R := V(G)\setminus X$ be the remaining independent set.

  In the first stage of the algorithm we guess, in time $O(2^{|X|})$, the intersection~$S \subseteq X$ of the maximal solution with $X$. If $S$ itself is not harmless, we discard it. Otherwise we create a modified instance $(G,t')$ where $t'(u) = t(u) - |N(u) \cap S|$, that is, we simply account for the budget used up by $S$. Since we have
  already guessed the intersection of the maximal solution and $X$, our goal is now to compute a maximal solution $I \subseteq R$ which is harmless in $(G,t')$. It is easy to verify that $I\cup S$ is then harmless in $(G,t)$.

  Note that finding a solution $I \subseteq R$ means that we can
  ignore the thresholds of vertices in $R$, only the thresholds of vertices in $X$
  constrain our solution. We proceed by partitioning~$R$ according
  neighbourhoods in $X$: for $A \subseteq X$, let $R_A$ contain all vertices
  $u \in R$ with $N(u) = A$. Since we can ignore thresholds of vertices in $R$,
  a solution $I \subseteq R$ can be encoded by simply noting the size of the
  intersection $x_A := |I \cap R_A|$ for all $A \subseteq X$.

  We will now formulate the problem as an ILP with at most $2^{2\vc(G)}$
  variables, that comprises the following parts
  \begin{enumerate}
  \item maximise the sum of chosen vertices
  \item each variable $x_A$, corresponding to the set whose
    neighbourhood is $A$, has size at most $|R_A|$
  \item each vertex $u \in X$, is at most its threshold $t'(u)$.
  \end{enumerate}
  Accordingly, the following
  ILP solves our subproblem:
  \begin{alignat*}{2}
    & \max & & \sum_{A \subseteq X} x_A \\
    & ~~\text{s.t.}  & \quad & 0 \leq x_A \leq |R_A| \quad\forall A \subseteq X \\
    &              &       & \! \sum_{\{u\} \subseteq A \subseteq X} \!\! x_A < t'(u) \quad\forall u \in X
  \end{alignat*}
  The first constraint ensures that our solution is realizable in $G$,
  while the second constraint ensures that it does not exceed the thresholds in $X$.
  This ILP has at most $2^{|X|}$ variables and we can therefore solve
  it in \FPT-time using Lenstra's algorithm~\cite{Lenstra83}. After solving all
  $2^{|X|}$ sub-problems, we return the largest total solution size (including the guessed intersection with $|X|$).
\end{proof}

\subsection{Modulator to $2$-spider-forest}

An instance of \name{Multicoloured Clique} consists of a $k$-partite graph
$G = (V_1,\ldots,V_k,E)$. The task is to find a clique
which intersects each colour $V_i$ in exactly one vertex. Since
\name{Multicoloured Clique} is \W[1]-hard~\cite{cygan2015parameterized}, our reduction establishes the
same for \HS{}.

In the following, we fix an instance $(V_1,\ldots,V_k,E)$ of \name{Multicoloured Clique}.
By a simple padding argument, we can assume that the sizes of the sets
$V_i$ are all the same and we will denote this cardinality by $n$ (thus the
graph has a total of $nk$ vertices). For convenience, we let $v^i_1,\ldots,v^i_n$
be the vertices of the set $V_i$. For indices $1 \leq i < j \leq k$ we denote
by $m_{ij} = |E(V_i, V_j)|$ the number of edges between colours $V_i$ and $V_j$.
We further let $m$ be the total number of edges.

\marginnote{remaining budget}
Finally, we will often speak of the remaining \emph{budget} of a vertex~$u$
with respect to some (partial) solution. This budget is to be understood
as the number of vertices in $N(u)$ that we can still select \emph{without}
violating the threshold $t(u)$. So if a partial solution has selected already
$s$ vertices in $N(u)$, then the remaining budget will be $t(u) - s - 1$.

\subsubsection*{Forbidden vertices}\hfill\\[0ex]
\noindent
Let $F \subseteq V(G)$ be a set of vertices that we want to prevent
from being in any solution. To that end, we construct a global
\emph{forbidden set} gadget which enforces that no vertex from $F$ can be selected.  The construction is similar to the \emph{forbidden}

\begin{wrapfigure}{l}{.5\textwidth}
\vspace*{-1em}
\includegraphics[width=.48\textwidth]{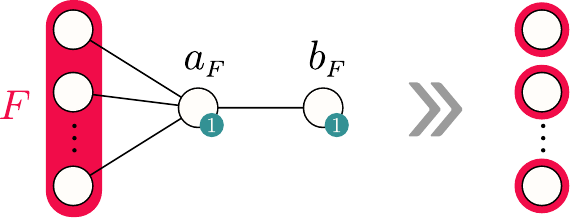}
\vspace*{-1em}
\end{wrapfigure}

\noindent
\emph{edge} gadget by Bazgan and
Chopin~\cite{HSParameterized14}:
We add two vertices $a_F$ and $b_F$ with threshold one to the graph and
make them connected.  Then we connect $a_F$ to every vertex in $F$.

\noindent
In the following gadgets we will often mark vertices as ``forbidden''.
We will denote this graphically by drawing a thick red border around
these vertices.

\begin{observation}
  Let $F,a_F,b_F$ be vertices as above in some instance $(G,t,k)$ of
  \HS{}. Then for every harmless set $S$ of $(G,t)$
  it holds that $S \cap (F \cup \{a_F,b_F\}) = \emptyset$.
\end{observation}

\subsubsection*{XOR gadget}\hfill\\[0ex]
\noindent
We construct an \emph{XOR gadget} for vertices $u$ and $v$ by adding a new forbidden
vertex~$x$ with threshold two and adding the edges $xu$ and $xv$ to 

\begin{wrapfigure}{r}{.5\textwidth}
\vspace*{-1em}%
\raggedleft\includegraphics[width=.48\textwidth]{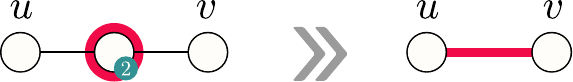}
\end{wrapfigure}

\noindent
the graph.
To simplify
the drawing of the following gadgets, we will simply draw a thick red edge
between to vertices to denote that they are connected by an XOR gadget.

\begin{observation}
  Let $u,x,v$ be as above in some instance $(G,t,k)$ of
  \HS{}. Then for every harmless set $S$ of $(G,t)$
  it holds that $|S \cap \{u,v\}| \leq 1$.
\end{observation}

\noindent
We will later enforce that in any solution $S$, $|S \cap \{u,v\}| = 1$,
hence the name XOR.

\vspace*{-1em}
\subsubsection*{Selection gadget}\hfill\\[1ex]
\begin{minipage}{.43\textwidth}
\raggedright\includegraphics[width=\textwidth]{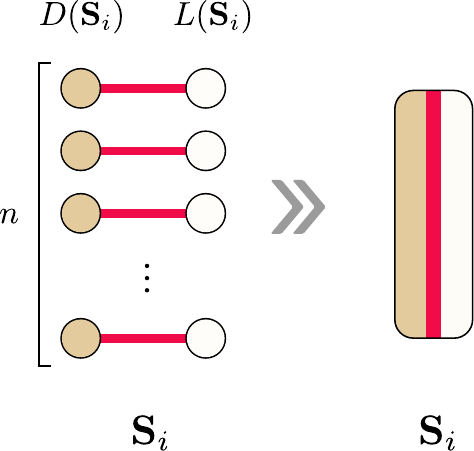}
\end{minipage}\hspace*{12pt}%
\begin{minipage}{.535\textwidth}
The role of a selection gadget $\mathbf S_i$ will be to select a single vertex from
one coloured set~$V_i$. The final construction will therefore contain $k$
of 
these gadgets $\mathbf S_1, \ldots, \mathbf S_k$. 
The gadget consists of $n$
pairs of vertices $d_s l_s$, $s \in [n]$,
 where each pair is connected
by an XOR gadget.
We call the set $D(\mathbf S_i) = \{d_1,\ldots,d_n\}$ the \emph{dark} vertices
and $L(\mathbf S_i) = \{l_1,\ldots,l_n\}$ the \emph{light} vertices.
We make two simple observations about the behaviour of this gadget:
\end{minipage}

\begin{observation}\label{obs:selectionA}
  Let $\mathbf S_i$ be as above in some instance $(G,t,k)$ of
  \HS{}. Then for every harmless set $S$ of $(G,t)$
  it holds that $|S \cap (D(\mathbf S_i) \cup L(\mathbf S_i)) | \leq n$.
\end{observation}

\noindent
By choosing an appropriate budget we will expect a solution to the final
instance to pick exactly $n$ vertices in each selection gadget and this
number encodes a vertex from the \name{Multicoloured Clique} instance. For
these solutions, we have that the number of vertices in the light and dark
part sum up exactly to $n$:

\begin{observation}\label{obs:selectionB}
  Let $\mathbf S_i$ be as above in some instance $(G,t,k)$ of
  \HS{}. Then for every harmless set $S$ of $(G,t)$
  with $|S \cap (D(\mathbf S_i) \cup L(\mathbf S_i)) | = n$ it holds
  that $|S \cap D(\mathbf S_i)| + |S \cap L(\mathbf S_i)| = n$.
\end{observation}

\subsubsection*{Port gadget}\hfill\\[0ex]

\begin{tightcenter}
\includegraphics[width=.8\textwidth]{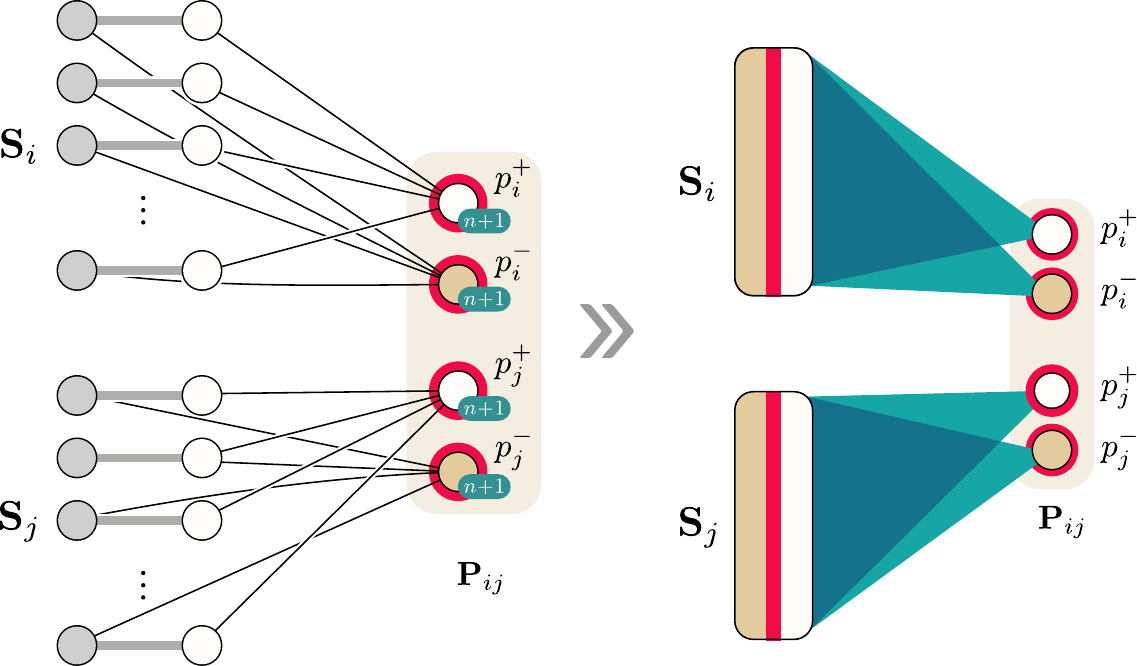}
\end{tightcenter}
\smallskip

\noindent
For every pair of selection gadgets $\mathbf S_i$, $\mathbf S_j$ we need to
communicate the choices these gadgets encode to further gadgets (described below)
which verify that this choice corresponds to an edge in $E(V_i,V_j)$.

The port gadget $\mathbf P_{ij}$ responsible for the pair $\mathbf S_i$, $\mathbf S_j$ consists of four forbidden \emph{port vertices} $p^+_i$, $p^-_i$, $p^+_j$, and $p^-_j$, each
with a threshold of $n+1$. For $\ell \in \{i,j\}$, we connect the
port vertex $p^+_\ell$ to the light vertices $L(\mathbf S_\ell)$ and the
port vertex $p^-_\ell$ to the dark vertices $D(\mathbf S_\ell)$. Note that
every selection gadget will be connected to $k-1$ port gadgets in this manner
and our naming scheme of the variables $p^+_{\any}$, $p^-_{\any}$ does not reflect
that. However, we will in the following only ever talk about a single port
gadget and therefore it will always be clear to which vertices we refer.

\subsubsection*{Test gadget}\hfill\\[0ex]
\noindent
The final gadget $\mathbf T_{xy}$ exists to test whether
two selection gadgets $\mathbf S_i$, $\mathbf S_j$ selected the edge $v^i_x v^j_y \in E(V_i,V_j)$. If that is the case, the gadget allows the inclusion of $n$ vertices into
the solution; otherwise it only allows the inclusion of a single vertex.

\smallskip
\begin{tightcenter}
\includegraphics[width=.7\textwidth]{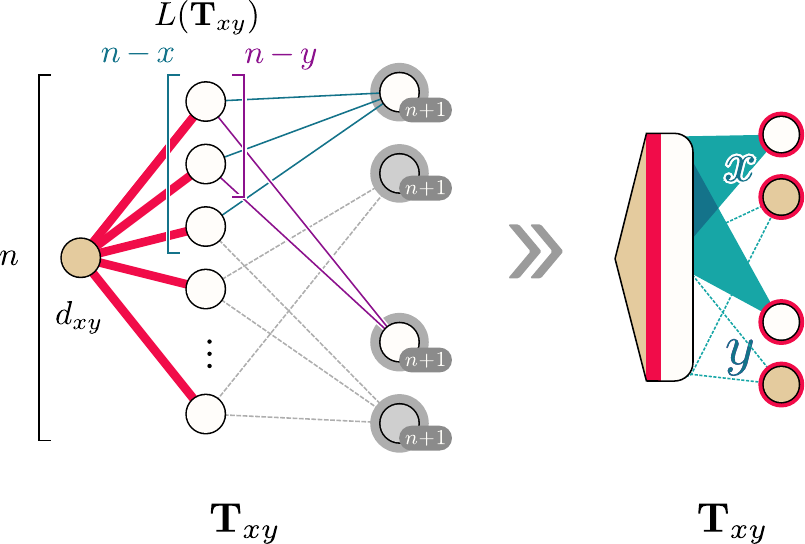}
\end{tightcenter}

The gadget consists of $n$ ordered light vertices $L(\mathbf T_{xy}) = \{ l_1,\ldots,l_n \}$ which are
all connected to a single dark vertex $d_{xy}$ via XOR gadgets. This already concludes the structure of
the gadget itself, but we need to discuss how it will be wired to the selection
gadgets $\mathbf S_i$ and $\mathbf S_j$ via the port gadget $\mathbf P_{ij}$.

For $i,j$ fixed as before, we connect the port $p^+_i \in \mathbf P_{ij}$
to the first $n-x$ light vertices
$l_1,\ldots,l_{n-x}$ and the port $p^-_i \in \mathbf P_{ij}$ to the last $x$ light
vertices $l_{n-x+1},\ldots,l_n$. Similarly, we connect the port $p^+_j \in \mathbf P_{ij}$
to the first $n-y$ light vertices
$l_1,\ldots,l_{n-y}$ and the port $p^-_j \in \mathbf P_{ij}$ to the last $y$ light
vertices $l_{n-y+1},\ldots,l_n$.

The idea of this construction is as follows:
If the selection gadget $\mathbf S_i$ ``selects'' the vertex $x$
and $\mathbf S_j$ ``selects'' $y$, our test gadget $\mathbf T_{xy}$ verifies
that the edge $xy$ exists in the original graph $G$ by allowing the inclusion
of all $n$ light vertices $L(\mathbf T_{xy})$. All other test gadgets $\mathbf T_{uv}$, $uv \neq xy$, wired to $\mathbf P_{ij}$ will, as we prove below, only allow the inclusion of their respective dark vertex $d_{uv}$.

\subsubsection*{Full construction}\hfill\\[0ex]
\noindent
The full construction for the reduction looks as follows. Given the instance
$G = (V_1 \uplus \cdots \uplus V_k, E)$ of \name{Multicoloured Clique}, we construct
an instance $(H,t)$ of \HS{} as follows:

\begin{itemize}
\item We add $k$ selection gadgets $\textbf S_1, \ldots, \textbf S_k$.
\item For every pair of indices $1 \leq i < j \leq k$:
\begin{itemize}
  \item We add the port gadget $\mathbf P_{ij}$ and connect it to $\mathbf S_i$ and $\mathbf S_j$ as described above.
  \item We add $m_{ij} := |E(V_i,V_j)|$ test gadgets $\{ \mathbf T_{xy} \}_{xy \in E(V_i,V_j)}$.
  \item We wire each test gadget $\mathbf T_{xy}$ to $\mathbf P_{ij}$ as described above.
  \item We add a forbidden vertex $a_{ij}$ to $H$ with threshold $n+1$ and connect it to all
        light vertices $\bigcup_{xy \in E(V_i,V_j)} L(\mathbf T_{xy})$.
\end{itemize}
\item Finally, we add the vertices $a_F$ and $b_F$ to $H$ and connect $a_F$
to all vertices marked as ``forbidden'' in the gadgets as well as to $b_F$.
\end{itemize}

\smallskip
\begin{tightcenter}
\includegraphics[width=.95\textwidth]{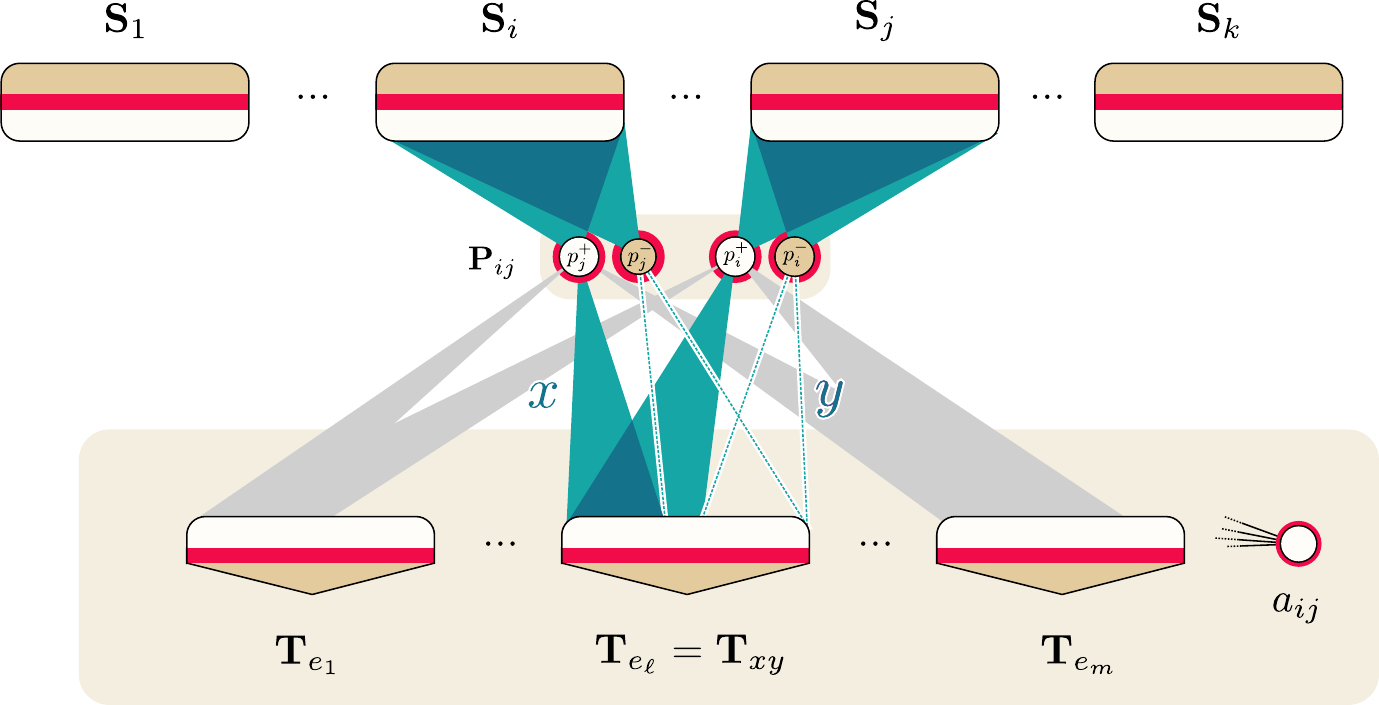}
\end{tightcenter}

\begin{lemma}\label{lemma:modulator}
  We can delete $5{k \choose 2} + 1$ vertices from $H$ to obtain a $2$-spider forest.
\end{lemma}
\begin{proof}
  We delete the $4{k \choose 2}$ vertices that make up the port gadgets,
  the ${k \choose 2}$ apices $a_{ij}$ for $1 \leq i < j \leq k$, and
  the vertex $a_F$. This disconnects all test- and selection gadgets from
  each other: the left-over vertices of the selection gadgets induce a forest
  of $P_3$s (the middle vertex being the XOR gadget vertex), while the left-over
  vertices of the test gadgets induce a $2$-spider forest.
\end{proof}

\def\Ssel{S_{\text{sel}}}
\def\Stest{S_{\text{test}}}
\begin{lemma}\label{lemma:completeness}
  If $G$ contains a multi-coloured clique on $k$ vertices, then $(H,t)$ has
  a harmless set of size ${k \choose 2}(n-1) + kn + m$.
\end{lemma}
\begin{proof}
  Let $x_1, \ldots, x_k$ be the indices of the clique-vertices, that is, the
  clique has vertices $v^i_{x_i}$ for $i \in [k]$. We construct a harmless set
  $S := \Ssel \cup \Stest $ as follows.

  First, let us construct $\Ssel$.  For each selection gadget $\mathbf S_i$,
  we select $x_i$ light vertices $l_1,\ldots,l_{x_i}$ from $L(\mathbf S_i)$
  and $n-x_i$ dark vertices $d_{x_i+1}, \dots, d_n$ from $D(\mathbf S_i)$.
  Observe that for each port gadget $\mathbf P_{i\any}$ (or $\mathbf P_{\any i}$), the remaining budget of $p^+_i$ is now $n-x_i$ and the remaining
  budget of $p^-_i$ is $x_i$. Note further that we did not include any forbidden
  vertices and the thresholds of the XOR gadgets have not been exceeded.

  Now, let us construct $\Stest$. As $v^1_{x_1}, \ldots, v^k_{x_k}$ induces a clique,
  we have that $v^i_{x_i}v^j_{x_j} \in E(G)$ for all $1 \leq i < j \leq k$. So for
  every pair of such indices $i,j$ we add all light vertices $L(\mathbf T_{x_ix_j})$
  of the test gadget $\mathbf T_{x_ix_j}$ to $\Stest$. For all remaining test gadgets $\mathbf T_{xy}$ with $xy \not \in \{ x_ix_j \mid 1 \leq i < j \leq k \}$
  we add the dark vertex $d_{xy}$ to $\Stest$.

  First, note that for every pair of indices $1 \leq i < j \leq k$ we selected
  exactly $n$ light vertices from all test gadgets wired to both $\mathbf S_i$ and $\mathbf S_j$.
  So in particular $|N(a_{ij}) \cap \Stest| = n$ and we therefore do not exceed the
  threshold of the apex $a_{ij}$. We also did not include any forbidden vertices and did not
  exceed the thresholds of the XOR gadgets inside the test gadgets as we either picked
  all light vertices (for $\mathbf T_{x_ix_j}$) or all dark vertices (all other test gadgets). Finally, consider the vertices $p^+_i$, $p^-_i$,
  $p^+_j$, and $p^-_j$ of the port gadget $\mathbf P_{ij}$. As observed above,
  the remaining budget after including $\Ssel$ of $p^+_\ell$ is $n - x_\ell$
  while the remaining budget of $p^-_\ell$ is $x_\ell$ for $\ell \in \{i,j\}$.
  By construction, $|N(p^+_\ell) \cap \Stest| = |N(p^+_\ell) \cap L(\mathbf T_{x_ix_j})|  = n-x_\ell$ and $|N(p^-_\ell) \cap \Stest| = |N(p^-_\ell) \cap L(\mathbf T_{x_ix_j})|  = x_\ell$ for $\ell \in \{i,j\}$, \ie $\Stest$ uses up exactly the budget left over by $\Ssel$.

  We conclude that the set $S = \Ssel \cup \Stest$ is indeed a harmless set of $(H,t)$. The total size of $S$ is
  \begin{align*}
    |S| &= |\Ssel| + |\Stest| = kn + \!\! \sum_{1 \leq i < j \leq k} \!\! (n+m_{ij}-1) \\
    &= kn + \!\! \sum_{1 \leq i < j \leq k} \!\! (n-1)  + \sum_{1 \leq i < j \leq k} \!\! m_{ij}\\
    &= kn + {k \choose 2}(n-1) + m,
  \end{align*}
  as claimed.
\end{proof}

\def\Lsel{L_\text{sel}}
\def\Dsel{D_\text{sel}}
\def\Ltest{L_\text{test}}
\def\Dtest{D_\text{test}}
\begin{lemma}\label{lemma:soundness}
  If $(H,t)$ has a harmless set of size ${k \choose 2}(n-1) + kn + m$, then $G$ contains a multi-coloured clique on $k$ vertices.
\end{lemma}
\begin{proof}
  Let $S$ be a harmless set of the above size. As we established above,
  $S$ cannot contain any vertices marked as ``forbidden'' in the construction.
  Therefore, $S$ can only contain light and dark vertices of the selection
  and test gadgets. Let us introduce the following shorthands:
  $\Lsel := \bigcup_{i \in [k]} L(\mathbf S_i)$ are the light vertices and $\Dsel := \bigcup_{i \in [k]} D(\mathbf S_i)$ the dark vertices inside selection gadgets. Similarly, let
  \begin{align*}
  \Ltest^{ij} &:= \!\bigcup_{xy \in E(V_i, V_j)}\! L(\mathbf T_{xy}) \quad\text{and} \\
  \Dtest^{ij} &:= \!\bigcup_{xy \in E(V_i, V_j)}\! D(\mathbf T_{xy}) = \{d_{xy} \mid xy \in xy \in E(V_i, V_j) \}.
  \end{align*}
  Let finally $\Ltest := \bigcup_{1 \leq i < j \leq k} \Ltest^{ij}$
  and $\Dtest := \bigcup_{1 \leq i < j \leq k} \Dtest^{ij}$ be the union of these
  sets.

  Let us now split up $S$ into $\Ssel := S \cap (\Lsel \cup \Dsel)$
  and $\Stest := S \cap (\Ltest \cup \Dtest)$. As all vertices outside of
  $\Lsel \cup \Dsel \cup \Ltest \cup \Dtest$ are forbidden, it follows that
  $\Ssel$ and $\Stest$ partition $S$. By Observation~\ref{obs:selectionA}
  we find that $|\Ssel| \leq k \cdot n$, as every selection gadget can contain
  at most $n$ vertices of $\Ssel$, and accordingly $|\Stest| \geq {k \choose 2}(n-1) + m$.

  To analyse the size and structure of $\Stest$, let us call a test gadget
  \emph{active} if $\Stest$ intersects its light vertices.

  \begin{claim}
    Fix an index pair $1 \leq i < j \leq k$. Let $\mathbf T_{e_1}, \ldots, \mathbf T_{e_s}$ be all active tests gadgets wired to $\mathbf P_{ij}$.
    Then $|\Stest \cap (\Ltest^{ij} \cup \Dtest^{ij})| \leq n + m_{ij} - s$
    if $s \geq 1$ and
    $|\Stest \cap (\Ltest^{ij} \cup \Dtest^{ij})| \leq m_{ij}$ otherwise.
  \end{claim}
  \begin{proof}[Proof of claim]
    Since $a_{ij}$ has a threshold of $n+1$, we know that $|\Stest \cap \Ltest^{ij}| \leq n$. Now note that, due to the XOR gadgets between the dark vertex
    and the light vertices of each test gadget, no dark vertex from $D(T_{e_1}) \cup \cdots
    \cup D(T_{e_s})$ can be contained in $\Stest$. Accordingly,
    $|\Stest \cap (\Ltest^{ij} \cup \Dtest^{ij})| \leq n + m_{ij} - s$. Consider now the
    case that $s = 0$, \ie there is no active gadget. Then $\Stest$ can
    only intersect the dark vertices $\Dtest^{ij}$ of which there are $m_{ij}$ many, accordingly  $|\Stest \cap \Dtest^{ij}| \leq m_{ij}$.
  \end{proof}

  \noindent
  Let $s_{ij}$ denote the number of active test gadgets attached to $\mathbf P_{ij}$. Then we can upper-bound the size of $\Stest$ by summing over the above bound:
  \begin{align*}
     |\Stest| &= \!\! \sum_{1 \leq i < j \leq k} \!\! |\Stest \cap (\Ltest^{ij} \cup \Dtest^{ij})| \\
     &\leq \!\! \sum_{1 \leq i < j \leq k} \!\! n\Iver{s_{ij} > 0} + m_{ij} - s_{ij} \\
    &= m + \!\! \sum_{1 \leq i < j \leq k} \!\! n\Iver{s_{ij} > 0}  - \!\! \sum_{1 \leq i < j \leq k} \!\! s_{ij}. \\
    &:= m + \nu n - \sigma.
  \end{align*}
  Where $\nu = \sum_{1 \leq i < j \leq k} \Iver{s_{ij} > 0}$ is the number of non-zero values $s_{ij}$ and  $\sigma :=  \sum_{1 \leq i < j \leq k} s_{ij}$
  is the sum of all $s_{ij}$. Note that $\sigma \geq \nu$.
  Comparing this upper bound and the previous lower
  bound on $\Stest$, we find that
  \[
    {k \choose 2}(n-1) + m \leq m + \nu n - \sigma
    \iff {k \choose 2}(n-1) \leq\nu n - \sigma.
  \]
  Since $\sigma \geq \nu$, we can weaken the above inequality to
  \[
    {k \choose 2}(n-1) \leq\nu n - \nu = \nu(n-1)
  \]
  from which we conclude that $\nu \geq {k \choose 2}$. Since $\nu > {k \choose 2}$
  is impossible, we have that $\nu = {k \choose 2}$. Therefore let us consider the updated inequality ${k \choose 2}(n-1) \leq {k \choose 2} n - \sigma$ which immediately implies that $\sigma \leq {k \choose 2}$. Since $\sigma \geq \nu = {k \choose 2}$ we find that
  $\sigma = \nu = {k \choose 2}$.

  Accordingly, the number of active gadgets
  is $s_{ij} = 1$ for \emph{all} indices $1 \leq i < j \leq k$. In other words,
  for every index pair $i,j$ there is exactly \emph{one} active test gadget.
  Further, we find that $|\Stest| = {k \choose 2}(n-1) + m$. Taking these two facts together, it follows that not only is there exactly one active test gadget
  per index pair, but $\Stest$ must contain \emph{all} of its $n$ light vertices.
  Let $\hat x_i\hat x_j$ for $1 \leq i < j \leq k$ be the indices of these active
  gadgets $\mathbf T_{\hat x_i\hat x_j}$.

  Having established the size and structure of $\Stest$, let us return to
  $\Ssel$. From the size of $\Stest$ we deduce that $|\Ssel| = kn$ and because $\Ssel$
  can intersect each selection gadget in at most $n$ vertices, it follows that
  $\Ssel$ intersects every selection gadget in exactly $n$ vertices. As noted in
  Observation~\ref{obs:selectionB}, this means that $|\Ssel \cap D(\mathbf S_i)|
  + |\Ssel \cap L(\mathbf S_i)| = n$ for all $i \in [k]$. Let $x_i :=
  |\Ssel \cap L(\mathbf S_i)|$, $i \in [k]$. Then for every port gadget $\mathbf P_{ij}$ it holds that $|N(p^+_\ell) \cap \Ssel| = x_\ell$ and $|N(p^-_\ell) \cap \Ssel| = n-x_\ell$ for $\ell \in \{i,j\}$.

  \begin{claim}
    Let $\hat x_i \hat x_j$ be the index of the active test gadget $\mathbf T_{\hat x_i \hat x_j}$ connected to the port $\mathbf P_{ij}$. Then $\hat x_i = x_i$
    and $\hat x_j = x_j$.
  \end{claim}
  \begin{proof}[Proof of claim]
    As established above, $\Stest$ contains all light vertices $L(\mathbf T_{\hat x_i \hat x_j})$. Consider the port vertices $p^+_\ell, p^-_\ell \in \mathbf P_{ij}$ for $\ell \in \{i,j\}$. Then
    \begin{align*}
      |N(p^+_\ell) \cap \Stest| &= |N(p^+_\ell) \cap L(\mathbf T_{\hat x_i \hat x_j})| = n - \hat x_\ell \\
      \text{and}\quad
      |N(p^-_\ell) \cap \Stest| &= |N(p^-_\ell) \cap L(\mathbf T_{\hat x_i \hat x_j})| = \hat x_\ell.
    \end{align*}
    On the other hand, we just established that
    $|N(p^+_\ell) \cap \Ssel| = x_\ell$ and $|N(p^-_\ell) \cap \Ssel| = n-x_\ell$.
    Accordingly,
    \begin{align*}
      |N(p^+_\ell) \cap S| &= |N(p^+_\ell) \cap \Ssel| + |N(p^+_\ell) \cap \Stest| = x_\ell + n - \hat x_\ell \\
      \text{and}\quad
      |N(p^-_\ell) \cap S| &= |N(p^-_\ell) \cap \Ssel| + |N(p^-_\ell) \cap \Stest| = n - x_\ell + \hat x_\ell
    \end{align*}
    As the threshold of $p^{\any}_\ell$ is $n+1$, we need that
    $n + x_\ell - \hat x_\ell \leq n$ and $n + \hat x_\ell - x_\ell \leq n$
    which of course only holds when $\hat x_\ell = x_\ell$.
  \end{proof}

  \noindent
  We therefore have that for all pairs of ``selected'' vertices $v^i_{x_i}, v^j_{x_j}$, $1 \leq i< j \leq k$, that the edge $v^i_{x_i}v^j_{x_j}$ exists in $G$ as witnessed by the existence of the (active) test gadget $\mathbf T_{x_ix_j}$. Accordingly, the $k$
  vertices $v^1_{x_1},\ldots,v^k_{x_k}$ form a multi-coloured clique in $G$, as claimed.

\end{proof}

\noindent
Lemma~\ref{lemma:modulator},~\ref{lemma:completeness}, and~\ref{lemma:soundness}
together prove Theorem~\ref{thm:lowerbound}.

\subsection{Subexponential time algorithm}

In order to apply the bidimensionality framework we will need to introduce
the following two annotated problems were we want solutions to avoid a certain
vertex subset.

\begin{XProblem}{Avoiding Harmless Set}
  \Input & A graph~$G$, an integer~$k$, a vertex set~$X \subseteq V(G)$. \\
  \Prob  & Does~$G$ have a harmless set~$S \subseteq V(G)\setminus X$ of size at least~$k$?
\end{XProblem}\vspace*{1ex}%
\begin{XProblem}{Avoiding $1$-Scattered Set}
  \Input & A graph~$G$, an integer~$k$, a vertex set~$X \subseteq V(G)$. \\
  \Prob  & Does~$G$ have a $1$-scattered set~$S \subseteq V(G)\setminus X$ of size at least~$k$?
\end{XProblem}

\noindent
In both cases, we call the vertices in~$X$ \emph{forbidden}. 
We say that a vertex is
\emph{simplicially forbidden} if it is forbidden and all its neighbors
are forbidden.  Observe that we may safely remove any simplicially
forbidden vertices for either of the two problems. We will assume in the following
that this preprocessing rule has been applied exhaustively and therefore
every forbidden vertex has at least one non-forbidden neighbour.

We define the contraction of an edge~$uv$ in an annotated graph $(G,X)$ as 
$(G,X) / uv := (G/uv, X')$ where
\[
  X' := \begin{cases*}
    (X \setminus \{u,v\}) \cup \{x_{uv}\} & if $\{u,v\} \subseteq X$ \\
     X \setminus \{u,v\}  & otherwise \\
  \end{cases*}
\]
and where~$x_{uv}$ is the vertex resulting from the contraction of~$u$ and~$v$.
In other words, the vertex~$x_{uv}$ is marked as forbidden iff both~$u$
and~$v$ were forbidden. A \emph{contraction minor} of $(G,X)$ is any annotated graph $(H,X')$ which can be obtained from $(G,X)$ by a sequence of contractions.

\begin{observation}
  \Problem{Avoiding $1$-Scattered Set} is closed under contractions, that is,
  if $(G,X)/uv$ has a solution of size~$k$ then so does~$(G,X)$.
\end{observation}
\begin{proof}
  Let~$S$ be a $1$-scattered set in~$(G,X)/uv := (G/uv, X')$. If~$x_{uv} \not \in S$,
  we are done since pairwise the pairwise distances of vertices in
  $G$ are at least as large as in~$G/uv$. Thus assume~$x_{uv} \in S$. Accordingly, $x_{uv} \not \in X'$ and therefore at least one of
  $u,v$ is not in $X$, wlog assume $u \not \in X$. Then~$(S \setminus x_{uv}) \cup \{ u \}$ is a $1$-scattered set in $G$ since~$\dist_G(u,y) \geq \dist_{G/uv}(x_{uv},y)$ for all $y \in V(G) \setminus \{u,v\}$. This set furthermore avoids
  $X$ and therefore is a solution for~$(G,X)$ of size~$|S|$, proving that 
  the problem is closed under contractions.
\end{proof}

\noindent
Finally, observe that if~$(G,X,k)$ is a YES-instance of \Problem{Avoiding 1-Scattered Set} then it is also a YES-instance of \Problem{Avoiding Harmless Set}. We are now ready to apply the bidimensionality framework.

\begin{figure}[t]
  \centering
  \includegraphics[scale=.4]{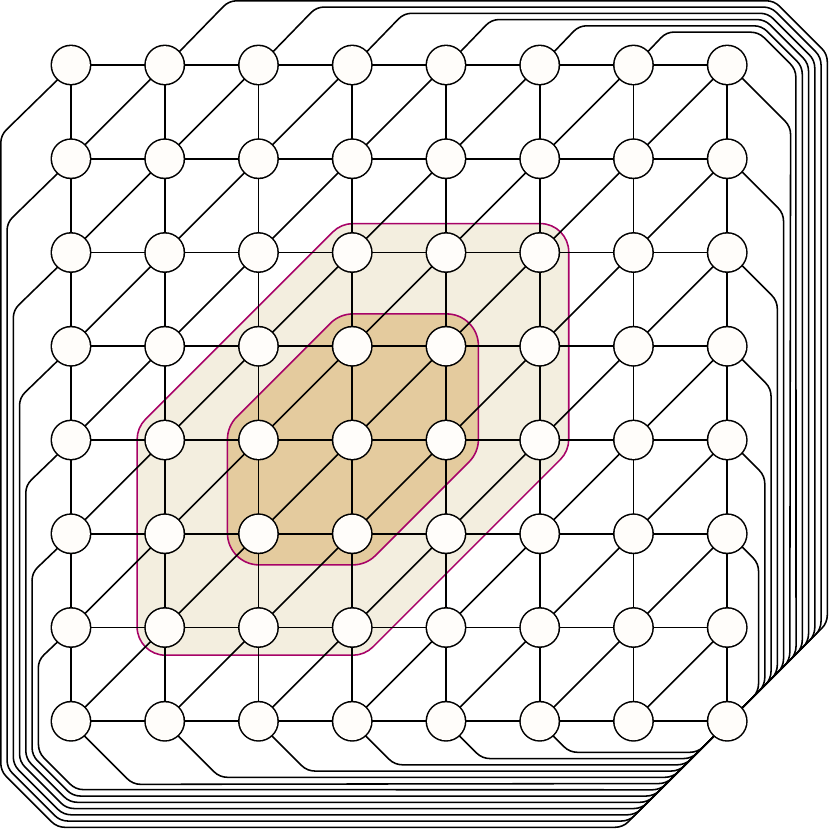}
  \caption{\label{fig:gamma}%
    The graph~$\Gamma_8$ with the first and second neighbourhood of an interior vertex marked.
  }
\end{figure}

\subexpalg*

\begin{proof}
  Fomin \etal~\cite[Theorem 1]{fomin2009contraction} proved 
  that for every apex-graph~$H$ there exists a constant $c_H$ such that if~$\tw(G) \geq k$ and~$G$ excludes~$H$ as a minor, then~$G$ has the graph~$\Gamma_{c_H\cdot k}$ as a contraction minor. Here~$\Gamma_t$ is the triangulated
  $t \times t$ grid where additionally one corner vertex is attached to all border vertices of the grid (\cf Figure~\ref{fig:gamma}).

  So assume that our input instance~$(G,X,k)$ has treewidth $\tw(G) \geq
  (5 \sqrt{k}+10) / c_H$, then $G$ contains $\Gamma_t$ as a contraction minor with~$t
  = 5\sqrt{k}+10$. Let $X' \subseteq V(\Gamma_t)$ be the contracted
  forbidden vertices as defined above. As we observed earlier, every vertex in
  $X'$ has at least one neighbour in~$\Gamma_t$ which is not in~$X'$.

  \begin{claim}
     $\Gamma_t$ contains a $1$-scattered set that avoids~$X'$ of size at least~$k$.
  \end{claim}
  \begin{proof}[Proof of claim]
    Assume that the vertices of $\Gamma_t$ are labelled $v_{i,j}$, where $i,j \in [t]$ denote the row-index and column-index of the respective vertex in the grid.

    Let~$S' := \{ v_{5x+3,5y+3} \mid 0 \leq x \leq (t-5)/5 ~\text{and}~ 0 \leq y \leq (t-5)/5 \}$. The set $S'$ is $2$-scattered in $\Gamma_t$ and
    has size at least~$(t/5-2)^2$. Every vertex $u \in S'$ is either not forbidden or it has a neighbour which is not forbidden, therefore we can construct a $1$-scattered set $S$ of the same size as follows:
    for every $u \in S'$ we add a non-forbidden vertex from $N[u]$ to $S$.
    The claim follows now since $S$ has size at least
    \[
      \big( \frac{t}{5} - 2 \big)^2 = 
      \big( \frac{5\sqrt{k}+10}{5} - 2 \big)^2 = k.
    \]
  \end{proof}
  \noindent 
  We conclude that if~$G$ has treewidth at least $w := (5 \sqrt{k}+10) / c_H$, then $(G, X, k)$ is a YES-instance. Using the single-exponential $5$-approximation for treewidth~\cite{bodlaender2016approximation}, we can in time $2^{O(w)} n = 2^{O(\sqrt{k})} n$ either find that $G$ has treewidth at least~$w$ or we obtain a tree decomposition of
  width no larger than $5w$. In the latter case, we use the algorithm
  by Bazgan and Chopin to solve the problem in time
  $k^{O(w)} n = 2^{O(\sqrt k \log k)} n$. Note that the total running time is bounded by $O(2^{o(k)} \cdot n)$, as claimed.
\end{proof}

\section{Conclusion}

We observed that the problem \HS{} is in \FPT{} for sparse graph classes due to existing machinery.
Therefore, we investigated its tractability in the
kernelization sense and found that \HS{} admits a polynomial \bekernel.
In the case of \name{$p$-Bounded Harmless Set} we even proved a
linear \bekernel.
We expect these results to extend to nowhere dense classes.

On the negative side, we demonstrated that sparseness alone does not make
the problem tractable. While the problem is in \FPT{} when parametrised by \eg
treewidth and solution size, we showed that it is in fact \W[1]-hard when
only parametrised by treewidth. Our reduction shows even more, namely that most sparse parameters (treedepth, pathwidth, feedback vertex set) can be ruled out as the problem is already hard when parametrised by a modulator to a $2$-spider-forest.

We conjecture---and leave as an interesting open problem---that \HS{} is already hard when parametrised by a modulator to a starforest.

\bibliographystyle{plain}
\bibliography{biblio}

\appendix
\section{Appendix}

\fptnowheredense*
\begin{proof}
  By Observation~\ref{obs:k-bounded}, \HS{} is equivalent
  to \name{$(k+1)$-Bounded Harmless Set}. Given an instance $(G,t,k)$ of
  the former, we can easily transform it into an instance $(G,t',k)$ of
  the latter where $t'(v) = \min\{t(v), k+1\}$.

  \newcommand{\phiharm}{\varphi^{HS}}
  We create the formula $\phiharm$ and prove that it defines
  \HS{}.

  Let $\psi(S, G, v)$ be the formula
  $\exists_{\leq t(v)} u . (v,u) \in E(G) \land u \in S$.
  Then
  $\phiharm_{k}(S, G) = ((|S| = k) \land \forall v . \psi(S, G, v))$.
  expresses that $S$ is a harmless set of size $k$. Note that
  the expressions $|S| = k$ and $\exists_{\leq t(v)}$ are both
  expressible in FOL, though the size of the resulting formula depends
  on $k$ and $\max_{v \in V} t(v) \leq k+1$.

  We now apply the powerful result by Grohe, Kreutzer, and Siebertz~\cite{NowhereDenseFO17}
  that a first-order sentence $\phi$ can be decided in time $O(n^{1+\epsilon})$
  for any $\epsilon > 0$ in nowhere dense classes. This algorithm is
  (non-uniformly) \FPT, concluding the proof.
\end{proof}

\lilylemma*
\begin{proof}
  Given $G$, we use Theorem~\ref{thm:dvorak-ds}  to compute a
  $d$-dominating set $D'$ of $A$ with~$|D'| = O(\ds_d(G,A))$ in
  polynomial time.  Afterwards, we compute the $(r+d)$-projection closure~$D$
  of $D'$, by Lemma~\ref{lemma:projclos} we have that $|D'| = O(|D|)$
  and therefore $|D| = O(\ds_d(G,A))$.
  Let $A'' := A \setminus D$, we will choose the polynomial $p_r$ so that
  $A''$ is still large enough for the following arguments to go through.

  Define the equivalence relation $\sim_D$ over $A''$ via
  \[
    a \sim_D a' \iff \pi^{r+d}_D[a] = \pi^{r+d}_D[a'].
  \]
  By Lemma~\ref{lemma:projbound}, the number
  of classes in $A'' / \sim_D$ is bounded by $O(|D|)$; by an averaging argument
  we have at least one class $[a] \in A'' / \sim_D$ of size
  \[
    \big|[a] \big| = \Omega\Big(\frac{|A''|}{|D|}\Big) = \Omega\Big(\frac{|A| - |D|}{|D|}\Big).
  \]
  Let $R'' = \proj^{r+d}_D(a)$, \eg
  the $(r+d)$-projection of $[a]$'s members onto $D$.
  By our earlier application of Lemma~\ref{lemma:projclos}
  we have that $|R''| = |\proj^{r+d}_D(a)| = O(1)$.

  We apply Theorem~\ref{thm:uqw} with distance~$r$ to the set $[a]$, let
  $g(r)$ be the function defined there. Using this notation, the algorithm
  of Theorem~\ref{thm:uqw} provides us, in polynomial time, with a subset
  $A' \subseteq [a]$ of size at least
  $
    |[a]|^{\frac{1}{g(r)}}
  $
  and a constant-sized set $R' \subseteq V(G)\setminus A'$,
  such that $A'$ is $r$-scattered in $G-R'$.

  Let $R := R' \cup R''$, by the above bounds on $R'$ and $R''$ it follows
  that $|R| = O(1)$.
  By Lemma~\ref{lemma:projbound} the number of different $d$-projections
  onto $|R|$ is bounded by $O(|R|)$, so we can find a set $C \subseteq A'$
  with uniform $d$-projections onto $|R|$ of size at least
  \[
    |C| \geq \frac{|A'|}{|R|} = \Omega(|A'|) = \Omega\Big(\big(\frac{|A| - |D|}{|D|}\big)^{\frac{1}{g(r)}}\Big).
  \]
  Since $|D| = O(\ds_d(G,A))$, there exists a polynomial
  $p_r(t) = O( (t^{1/g(r)}+1))$ such that $|A| \geq (t^{1/g(r)} + 1) \cdot |D|$, which implies that
  \[
    |C| = \Omega\Big(\big(\frac{|A| - |D|}{|D|}\big)^{\frac{1}{g(r)}}\Big)
    = \Omega(t).
  \]
  Therefore we can choose $p_r(t)$ so that $|C| \geq t$, as claimed.
\end{proof}

\end{document}

%% file: header.tex
\usepackage{bbm}
\usepackage{mathrsfs}
\usepackage{latexsym}
\usepackage{paralist}
\usepackage{microtype}

\usepackage{suffix}

\usepackage{xpunctuate}
\usepackage[all,british]{foreign} 
\redefnotforeign[ie]{i.e\xperiodafter} 
\redefnotforeign[eg]{e.g\xperiodafter}

\usepackage{xfrac} 
\usepackage[style=english,english=british]{csquotes}

\usepackage{amsmath, amssymb, amsfonts} 

\usepackage{amsthm}

\usepackage{mathtools}

\usepackage[full,small]{complexity}

\usepackage[final,notref,color]{showkeys} 

\usepackage{xargs}
\usepackage{afterpage} 

\PassOptionsToPackage{dvipsnames,usenames,table}{xcolor}
\usepackage{tikz}
\usetikzlibrary{calc}

\usepackage{booktabs} 
\usepackage{longtable}
\usepackage{float}
\usepackage{wrapfig}
\usepackage{floatflt}
\usepackage{framed}
\usepackage{ctable}
\usepackage{dcolumn}
\usepackage{adjustbox}

\usepackage[shortlabels]{enumitem} 
\usepackage{xspace}
\usepackage{xifthen}

\usepackage{url}
\usepackage{scrtime}

\usepackage[longend,vlined]{algorithm2e}

\usepackage{marginnote}

\makeatletter
\newcommand{\raisemath}[1]{\mathpalette{\raisem@th{#1}}}
\newcommand{\raisem@th}[3]{\raisebox{#1}{$#2#3$}}
\makeatother

\def\N{\mathbb{N}} 

\renewcommand{\cal}{\mathcal}

\newcommand{\name}[1]{\textnormal{\textsc{#1}}}

\newcommand\Problem[2][]{%
    \textsc{#2}\xspace%
}

\newcommand{\HS}{\Problem{Harmless Set}}

\newcommand{\pBHS}{\Problem{$p$-Bounded Harmless Set}}

\newclass{\paraNP}{paraNP}

\newcommand\restr[2]{{
  \left.\kern-\nulldelimiterspace 
  #1 
  \vphantom{\big|} 
  \right|_{#2} 
  }}

\def\sminor^#1{
    \preccurlyeq_{\mathrlap{\mathsf{m}}}^{#1}
} 
\def\ssminor^#1{%
    \mathbin{\dot\preccurlyeq_{{\mathrlap{\mathsf{m}}}}^{#1}}\,
} 
\def\stminor^#1{
    \preccurlyeq_{\mathrlap{\mathsf{t}}}^{#1}
} 
\def\sstminor^#1{%
    \mathbin{\dot\preccurlyeq_{{\mathrlap{\mathsf{t}}}}^{#1}}\,
} 

\def\grad_#1{\nabla\!_{#1}}
\def\sgrad_#1{{\dot\nabla}\!_{#1}}
\def\topgrad_#1{\widetilde \nabla\!_{#1}}
\def\stopgrad_#1{\dot{\widetilde\nabla}\!_{#1}}
\def\topomega_#1{\widetilde \omega_{#1}}

\newcommand{\dist}{\ensuremath{\text{dist}}}

\def\colnum_#1{ \operatorname{col}_{#1} }
\def\wcolnum_#1{ \operatorname{wcol}_{#1} }
\def\adm_#1{ \operatorname{adm}_{#1} }

\renewcommand{\leq}{\leqslant}

\renewcommand{\geq}{\geqslant}

\renewcommand{\epsilon}{\varepsilon}

\renewcommand{\emptyset}{\varnothing}

\newlength{\convarrowwidth}
\usepackage{stmaryrd}
\def\Iver#1{ \llbracket #1 \rrbracket }

\newcommand{\widthm}[1]{ \mathbf{#1} } 
\DeclareMathOperator{\tw}{ \widthm{tw} }

\DeclareMathOperator{\width}{ \widthm{width} } 

\def\YYYY{{Y_0 \uplus Y_1 \uplus \cdots \uplus Y_\ell}} 
\WithSuffix\def\YYYY'{{Y'_0 \uplus Y'_1 \uplus \cdots \uplus Y'_{\ell'}}}

\def\ds{\operatorname{\mathbf{dom}}}

\def\vc{\operatorname{\mathbf{vc}}}

\usepackage{pifont}
\newcommand{\yaay}{\kern4pt \ding{51} \kern-8pt \ding{51}}%

\def\any{\mathord{\color{black!33}\bullet}}%
\newtheorem{observation}{Observation}

\newtheorem*{open*}{Open question}
\usepackage{thmtools, thm-restate}

\renewcommand*\etal{\xperiodafter{\emph{et~al}}}

\newcommand*\varrule[1][0.4pt]{\leavevmode\leaders\hrule height#1\hfill\kern0pt}

\setlist[1]{labelindent=\parindent,leftmargin=*} 
\setlist{itemsep=0pt}
\setitemize[1]{label={\small\textbullet}}

\newenvironment{tightcenter}
 {\parskip=0pt\par\nopagebreak\centering}
 {\par\noindent\ignorespacesafterend}

\usepackage{ctable}
\newlength{\RoundedBoxWidth}
\newsavebox{\GrayRoundedBox}
\newenvironment{GrayBox}[1]%
   {\setlength{\RoundedBoxWidth}{\textwidth-4.5ex}
    \def\boxheading{#1}
    \begin{lrbox}{\GrayRoundedBox}
       \begin{minipage}{\RoundedBoxWidth}%
   }{%
       \end{minipage}
    \end{lrbox}%
    \begin{tightcenter}%
    \begin{tikzpicture}%
       \node(Text)[draw=black!20,fill=white,rounded corners,%
             inner sep=2ex,text width=\RoundedBoxWidth]%
             {\usebox{\GrayRoundedBox}};
        \coordinate(x) at (current bounding box.north west);
        \node [draw=white,rectangle,inner sep=3pt,anchor=north west,fill=white] 
        at ($(x)+(6pt,.75em)$) {\boxheading};
    \end{tikzpicture}
    \end{tightcenter}\vspace{0pt}%
    \ignorespacesafterend
}    

\newenvironment{XProblem}[2][]{\noindent\ignorespaces%
                                \FrameSep=6pt%
                                \parindent=0pt%
                \vspace*{-.5em}
                \ifthenelse{\isempty{#1}}{%
                  \begin{GrayBox}{\textsc{#2}}%
                }{%
                  \begin{GrayBox}{\textsc{#2} parametrised by~{#1}}%
                }
                \newcommand\Prob{Problem:}%
                \newcommand\Input{Input:}%
                \begin{tabular*}{\textwidth}{@{\hspace{.1em}} >{\itshape} p{1.6cm} p{0.8\textwidth} @{}}%
            }{
                \end{tabular*}%
                \end{GrayBox}%
                \vspace*{-.5em}
                \ignorespacesafterend
            }  
\newenvironment{XProblem*}[2][]{\noindent\ignorespaces%
                                \FrameSep=6pt%
                                \parindent=0pt%
                \vspace*{-.5em}
                \ifthenelse{\isempty{#1}}{%
                  \begin{GrayBox}{\textsc{#2}}%
                }{%
                  \begin{GrayBox}{\textsc{#2} parametrised by~{#1}}%
                }
                \newcommand\Prob{Problem:}%
                \newcommand\Input{Input:}%
                \begin{tabular*}{\textwidth}{@{\hspace{.1em}} >{\itshape} p{1.6cm} p{0.8\textwidth} @{}}%
            }{
                \end{tabular*}%
                \end{GrayBox}%
                \vspace*{-.5em}
                \ignorespacesafterend
            }       

\renewenvironmentx{leftbar}[2][1=2pt, 2=5pt]{%
  \def\FrameCommand{\vrule width #1 \hspace{#2}}\MakeFramed {\advance\hsize-\width \FrameRestore}}%
{\endMakeFramed}

\newlength{\wleft}  \newlength{\wright}

\definecolor{Maroon}{cmyk}{0, 0.87, 0.68, 0.32}
\definecolor{RoyalBlue}{cmyk}{1, 0.50, 0, 0}
\definecolor{Black}{cmyk}{0, 0, 0, 0}
\definecolor{White}{rgb}{1, 1, 1}
\definecolor{Red}{rgb}{.9, 0, .1}
\definecolor{Magenta}{rgb}{.7, 0, .7}

\makeatletter
\let\orgdescriptionlabel\descriptionlabel
\def\@savelabel{}
\renewcommand*{\descriptionlabel}[1]{%
  \let\orglabel\label
  \let\label\@gobble
  \phantomsection
  \def\@savelabel{#1}
  \edef\@currentlabel{{\def\hfil{}#1}}
  \edef\@currentlabelname{#1}%
  \let\label\orglabel
  \orgdescriptionlabel{#1}%
}
\makeatother
\makeatletter
\def\namedlabel#1#2{\begingroup
   \def\@currentlabel{#1}%
   \label{#2}\endgroup
}
\makeatother

\usepackage{hyphenat}
\renewcommand{\th}{%
    \ifmmode
        ^\mathrm{th}%
    \else%
        \textsuperscript{th}\xspace%
    \fi%
}
\newcommand{\st}{%
    \ifmmode
        ^\mathrm{st}%
    \else%
        \textsuperscript{st}\xspace%
    \fi%
}
\newcommand{\nd}{%
    \ifmmode
        ^\mathrm{nd}%
    \else%
        \textsuperscript{nd}\xspace%
    \fi%
}
\newcommand{\rd}{%
    \ifmmode
        ^\mathrm{rd}%
    \else%
        \textsuperscript{rd}\xspace%
    \fi%
}

\def\Nesetril{Ne\v{s}et\v{r}il\xspace}

\def\Dvorak{Dvo\v{r}\'{a}k\xspace}

\usepackage{environ}%
\NewEnviron{nofiitable}{\noindent\ignorespaces%
  
  \[\arraycolsep=1.4pt%
  \begin{array}{rcr p{1cm} rcr}
    \BODY
  \end{array}
  \]
}

\newcolumntype{m}{>{$}l<{$}}
\newcolumntype{M}{>{$\displaystyle}l<{$}} 

\newcolumntype{L}{l}  
\newcolumntype{C}{c}  
\newcolumntype{R}{r}  
\newcolumntype{X}{>{\global\let\currentrowstyle\relax}}
\newcolumntype{^}{>{\currentrowstyle}}